\documentclass{emulateapj}

\usepackage{apjfonts}

\begin{document}

\newcommand{\xc}{{x^{(2)}}}

\title{Three-Dimensional Simulations of Magnetized Thin Accretion Disks around Black Holes: Stress in the Plunging Region}

\author{Rebecca Shafee\altaffilmark{1},
Jonathan C. McKinney\altaffilmark{2},
Ramesh Narayan\altaffilmark{3},
Alexander Tchekhovskoy\altaffilmark{3}\\
Charles F. Gammie\altaffilmark{4},
Jeffrey E. McClintock\altaffilmark{3}}
\altaffiltext{1}{Harvard University, Department of Physics,
17 Oxford  Street, Cambridge, MA 02138}
\altaffiltext{2}{Chandra Fellow; Kavli Institute for Particle Astrophysics and Cosmology,
Stanford University, CA 94309}
\altaffiltext{3}{Harvard-Smithsonian Center for Astrophysics, 60 Garden
Street, Cambridge, MA 02138}
\altaffiltext{4}{Center for Theoretical Astrophysics, University
of Illinois, Urbana-Champaign, IL 61801}

\begin{abstract}

We describe three-dimensional general relativistic magnetohydrodynamic
simulations of a geometrically thin accretion disk around a
non-spinning black hole. The disk has a thickness $h/r\sim0.05-0.1$
over the radial range $(2-20)GM/c^2$.  In steady state, the specific
angular momentum profile of the inflowing magnetized gas deviates by
less than 2\% from that of the standard thin disk model of
\citet{nt73}.  Also, the magnetic torque at the radius of the
innermost stable circular orbit (ISCO) is only $\sim2\%$ of the inward
flux of angular momentum at this radius.  Both results indicate that
magnetic coupling across the ISCO is relatively unimportant for
geometrically thin disks.

\end{abstract}

\keywords{X-ray: stars --- binaries: close --- accretion, accretion
disks --- black hole physics}

\section{Introduction}

The recent development of general relativistic magnetohydrodynamic
(GRMHD) codes (e.g., \citealt{gmt03,dev03}) has finally allowed
realistic numerical simulations of magnetized accretion disks around
black holes (BHs).  This has led to a better understanding of the
inner regions of these disks (e.g. \citealt{khh05}) and of their role
in launching relativistic jets (e.g. \citealt{M06}).  One of the
interesting new results is the recognition that magnetic fields alter
the structure of the accretion flow near and inside the innermost
stable circular orbit (ISCO).  This may result in large deviations
from the traditional picture of a vanishing torque at the ISCO
\citep{krolik99,gammie99,khh05}.

\citet{pac00} and \citet{ap03} suggested that the zero-torque
condition is likely to be a good approximation for geometrically thin
disks.  This was confirmed by \citet{snm08} who, using a global
height-integrated model, showed that modifications to the stress
profile are negligibly small for disk thicknesses $h$ less than about
a tenth of the local radius $r$.  Their work was, however, based on a
hydrodynamic model with $\alpha$-viscosity and did not explicitly
include magnetic fields.

Numerical MHD simulations by \citet{kh02} using a pseudo-Newtonian
potential, and by \citet{khh05} using a GRMHD code, indicated that
magnetic torques are indeed important inside the ISCO.  In particular,
these authors found no evidence for a ``stress edge,'' leading them to
argue that simple disk models based on the zero-torque condition
(e.g., \citealt{ss73}; \citealt{nt73}, hereafter NT73) may be
seriously wrong. If true this would undermine recent efforts to
estimate of the spin parameters of BHs using the NT73 model
\citep{shafee06, mcc06, ddb06,liu08}.

The MHD simulations carried out so far have
considered non-radiating accretion flows that are geometrically
rather thick.  The one exception is the recent work of \citet{rb08}
who considered a thin disk with $h/r\sim 0.05$ in a
pseudo-Newtonian potential.  In this paper, we describe global 3D GRMHD
simulations of a thin disk ($h/r\sim0.05$) around a non-spinning
BH and compare our simulated model with the NT73 model.

\section{Numerical Model}

We use units with $G=c=1$, e.g. the horizon is at $r=2M$ and the ISCO is
at $6M$.  We report results in Boyer-Lindquist (BL) coordinates ($t$,
$r$, $\theta$, $\phi$), referred to as the coordinate frame.  In
the expressions below $g$ refers to the determinant of the metric. The
fluid $4$-velocity and the magnetic field $4$-vector (see, e.g.,
\citealt{anile89} for definitions) are given by $u^\mu$ and
$b^{\mu}$, and the rest-mass density, internal energy density, thermal
pressure, and magnetic pressure as measured in the fluid comoving
frame are $\rho$, $u_g$, $p=(\gamma-1)u_g$ with adiabatic index
$\gamma=4/3$, and $p_b=b^2/2$.  The total pressure is $p_{\rm tot} = p
+ p_b$.

Simulations were performed using a 3D GRMHD code HARM
\citep{gmt03} in Kerr-Schild coordinates using the interpolation
scheme described by \citet{M06}, inversion scheme described
by \citet{N06} and \citet{mm07}, and other advances described by \citet{tmn07}.  For
our fiducial model we used a 3D grid with resolution $512\times
128\times 32$ corresponding to: (i) 512 cells in $r$, logarithmically
spaced from $r=1.8M$ to $50M$ with ``outflow'' boundary conditions;
(ii) 128 cells in $\theta$ going from 0 to $\pi$, non-uniformly spaced
so that roughly half the cells are concentrated in the
disk\footnote{We used a grid given by $\theta(\xc) = [h(2\xc-1) +
(1-h)(2\xc-1)^7 +1]\pi/2$ for code coordinate $\xc$ with
$h=0.15$, giving roughly $6\times$ more angular resolution compared
to equation~(8) with $h=0.3$ in \citet{mg04}.} ; (iii) 32 cells in
$\phi$, uniformly spaced from $0$ to $\pi/4$.  Cells at the disk
equator had physical sizes roughly in the ratio of 2:1:7 in $dr$,
$r d\theta$, $r\sin(\theta) d\phi$, which ensured that the turbulence was roughly
isotropic in Cartesian coordinates, optimal for
accurately resolving the magnetorotational instability (MRI).  To test
convergence we also used resolutions of $256\times64\times16$,
$256\times64\times32$, and $256\times64\times64$.  We also carried out
2D simulations with resolution of up to $2048\times256\times1$,
but we do not discuss these results because turbulence decayed
on an orbital time-scale (i.e. Cowling's anti-dynamo theorem holds).

We began the simulation with an equilibrium torus \citep{chak85,dev03} with
inner edge at $r=20M$ and pressure maximum at $r=35M$ and adjusted the
model parameters so the torus had $h/r=0.1$ at $35M$.  We found that
placing the torus at a smaller radius led to results too sensitive to
the initial mass distribution.  We define $h/r$ to be the
density-weighted root mean square angular thickness of the disk at any
given $r$, i.e.
\begin{equation}
\label{thickness}
\left({h\over r}\right)_r = \Delta\theta_{\rm rms} = \left[\frac{\int\int
(\Delta\theta)^2\rho(r,\theta,\phi) \sqrt{-g} d\theta
d\phi}{\int\int \rho(r,\theta,\phi) \sqrt{-g} d\theta
d\phi}\right]^{1/2} ,
\label{hoverr}
\end{equation}
or written as $\Delta\theta_{\rm rms} = \langle \Delta\theta^2 \rangle^{1/2}$
with $\Delta\theta\equiv\theta-\langle\theta\rangle$.
We also consider the mean density-weighted thickness: $h/r =
{\Delta\theta}_{\rm abs}=\langle|\Delta\theta|\rangle$.

We embedded the torus with a weak magnetic field with
$\beta\equiv p/p_b\sim 100$.  The initial field consisted of two poloidal
loops centered at $r=28M$ and $38M$ to model a disorganized field
with no net flux.  (During the simulation, there is no
organized flux threading the disk but some organized flux threads the BH.)
The field strength was randomly perturbed by 50\% to seed the MRI instability.
Recent GRMHD simulations \citep{mn07a,mn07b,beckwith08a} indicate that the results for the disk
(but not the jet) should be roughly independent of the initial field
geometry.  The MRI is initially resolved with much of
the torus having $10$ cells per wavelength of the fastest growing MRI
mode.

In order to keep the accretion disk thin, an ad hoc cooling/heating
function was added to the energy-momentum equations as a covariant
source term ($-u^\mu\,du_g/d\tau$) with
\begin{equation}
{du_g\over d\tau} = -{u_g-u_{\rm eq}\over \tau_{\rm cool}},
\end{equation}
where $\tau$ is the fluid proper time.  The gas cooling time
($\tau_{\rm cool}$) was set to $2\pi/\Omega_K$, where
$\Omega_K=(r/M)^{-3/2}M$ is the Keplerian frequency.  Thus the gas was
driven towards $u_{\rm eq}$, which we defined as that value of $u_g$
for which the specific entropy of the gas would be equal to the
constant specific entropy (e.g., $p/\rho^\gamma$) of the initial
solution.

\begin{figure}
\includegraphics[width=3.3in,clip]{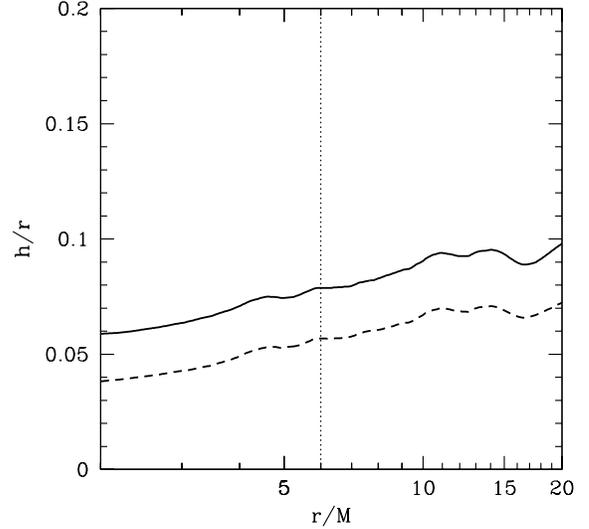} 
\caption{Variation of the root mean square disk thickness
$h/r=\Delta\theta_{\rm rms}$ (eq. \ref{thickness}) as a function of
radius $r$ (solid line).  Simulation results were averaged over the
time interval $6000M-10000M$.  Also shown is the mean absolute
thickness ${\Delta\theta}_{\rm abs}$ (dashed line).  The vertical dotted line
corresponds to the position of the ISCO.  The initial torus had its
inner edge at $r_{\rm in}=20M$.}
\label{thicknessplot} \end{figure}

The simulation ran for a time of $10000M$, corresponding to $108$
orbits at the ISCO ($r_{\rm ISCO}=6M$) and $18$ orbits at the initial
inner edge of the torus. The results reported here correspond to
averages computed over the period $6000M-10000M$, when the accretion
flow had reached a quasi-steady state inside a radius of about $10M$.
Figure~\ref{thicknessplot} shows that the simulated disk had a
thickness (given by eq. \ref{thickness}) of $h/r\sim0.06-0.10$ over
the radius range of interest.  The mean absolute thickness
${\Delta\theta}_{\rm abs}$ was smaller: $\sim0.04-0.07$.  At the ISCO, the two
definitions of thickness gave $h/r\sim0.08, ~0.06$, respectively.

\section{Results}

The flux of mass and specific angular momentum are given by
\begin{eqnarray}
\dot{M}(r,t) &=& -\int\int \rho u^r \sqrt{-g} d\theta d\phi, \\
{\dot{L}(r,t)\over\dot{M}(r,t)} &=& -{1\over\dot{M}(r,t)}\int\int
\left(T_{\rm \phi, in}^r - T_{\rm \phi, out}^r \right)\sqrt{-g}d\theta
d\phi \\
&\equiv& \bar{\ell}_{\rm in}(r,t) -\bar{\ell}_{\rm out}(r,t),
\label{Ldot}
\end{eqnarray}
respectively, where  $T^{r}_{\phi}$ is the $r$-$\phi$ component
of the stress-energy tensor, and
\begin{equation}
T_{\rm \phi, in}^r = (\rho+u_g+p+b^2)u^r u_\phi, \qquad
T_{\rm \phi, out}^r = b^r b_\phi.
\end{equation}
The specific energy flux $\dot{E}/\dot{M}$ is found by replacing
$T^r_\phi$ with $T^r_t$ in the above equations.  In equation
(\ref{Ldot}), the term $\bar{\ell}_{\rm in}$ represents the specific
angular momentum advected inward with the accretion flow.  Similarly,
$\bar{\ell}_{\rm out}$ represents outflow of specific angular momentum
as a result of the magnetic shear stress.  In steady state, each of
these terms is independent of $t$, and their sum is independent of
both $r$ and $t$.

\begin{figure}
\includegraphics[width=3.3in,clip]{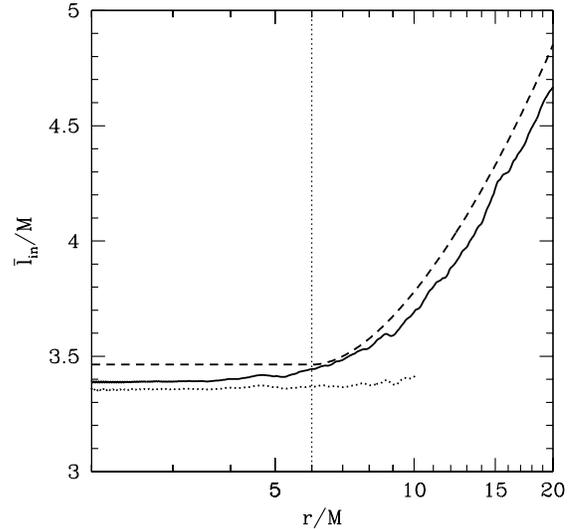} 
\caption{Time-averaged profile of the specific angular momentum
$\bar{\ell}_{\rm in}$ of the inflowing magnetized gas (solid line),
compared to the idealized profile assumed in the NT73 disk model
(dashed line).  The horizontal dotted curve shows the net flux of
angular momentum $\bar{\ell}_{\rm in}-\bar{\ell}_{\rm out}$.  This is
independent of $r$ up to $\sim 10M$, indicating the simulation has
achieved steady state well-beyond the ISCO.}
\label{angmmtm} \end{figure}

Figure~\ref{angmmtm} compares the time-averaged profile of
$\bar{\ell}_{\rm in}$ from the simulation, averaged over an angular
range $\delta\theta=\pm0.2$ around the disk mid-plane (2 to 3 density
scale heights), with that predicted by NT73.  In the latter,
$\bar{\ell}_{\rm in}$ is equal to the Keplerian specific angular
momentum for all radii down to the ISCO; inside the ISCO,
$\bar{\ell}_{\rm in}$ is taken to be constant since, by assumption, no
angular momentum is removed from the gas in the plunging region.  The
$\bar{\ell}_{\rm in}$ profile from the simulation shows modest
deviations from this idealized profile: (i) $\bar{\ell}_{\rm in}$ is
slightly sub-Keplerian outside the ISCO; (ii) $\bar{\ell}_{\rm in}$
continues to decrease for a range of $r$ inside the ISCO; (iii)
$\bar{\ell}_{\rm in}$ becomes essentially independent of $r$ close to
the horizon.

The surprising feature of Fig. 2 is that the two curves are so close
to each other inside the ISCO.  While it is true that the profile of
$\bar{\ell}_{\rm in}$ from the simulation drops inside the ISCO to a
value of $3.39M$, this value is only 2.0\% less than the NT73 value of
$3.464M$.  Thus, the magnetic coupling that technically operates
inside $r_{\rm ISCO}$ does not have much real effect on the accreting
gas.  At the ISCO $\bar{\ell}_{\rm out}=0.075M$, which is $\sim2.2\%$
of $\bar{\ell}_{\rm in}$.  Further, $\dot{E}/\dot{M}$ at the horizon
is $0.940$, which is only slightly different from the NT73 value of
$0.9428$, showing that the accretion efficiency is not enhanced by
magnetic fields.  For two-dimensional simulations of thicker disks
with $h/r\sim 0.3$, \citet{mg04} found that $\dot{L}/\dot{M} = 3.068$
and $\dot{E}/\dot{M} = 0.950$.  It appears that $\dot{L}/\dot{M}$
deviates from the NT73 value roughly proportional to $h/r$ and so
thinner disks should be even closer to NT73.  These results
are converged since, between a resolution of $256\times 64\times 32$
and $512\times 128\times 32$, the value of $\dot{L}/\dot{M}$ changes
by less than $0.2\%$ and $\dot{E}/\dot{M}$ changes by less than
$0.5\%$.

\begin{figure}
\includegraphics[width=3.3in,clip]{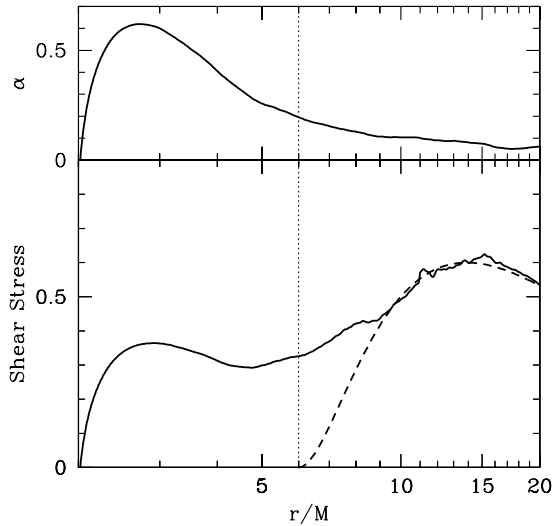} 
\caption{Lower panel shows time-averaged profile of the fluid shear stress (solid line)
compared to the NT73 stress (dashed line).  Upper panel shows $\alpha$ viscosity coefficient (solid line).
The non-zero stress in the plunging region causes only a minor deviation from the NT73 result for
the flux of specific angular momentum.}
\label{fluidstress} \end{figure}

Figure~\ref{fluidstress} shows the time-average of the normalized shear stress
\begin{equation}
\tilde{W} \equiv 10^3 \int\int T_{\hat{\phi}\hat{r}} \sqrt{-g} d\theta d\phi / (2\pi r \dot{M}) ,
\end{equation}
where $W\equiv 10^{-3} \tilde{W}\dot{M}$ as equation (5.6.1b) in NT73 and
$T_{\hat{\phi}\hat{r}}$ is the orthonormal stress-energy tensor
components in the comoving frame.
We restrict the integral to fluid with $u_t
(\rho+u+p+b^2)/\rho > -1$, i.e., only gravitationally bound
fluid (no disk wind); this lowers the value of $W$ near the horizon by
50\%.
Figure~\ref{fluidstress} also shows the space-time averaged
viscosity parameter $\alpha\equiv T_{\hat{\phi}\hat{r}}/p_{\rm tot}$
that is roughly $0.1-0.15$ outside the ISCO.  Between the ISCO and
$r=3M$, $\alpha$ rises to $0.6$ and finally drops to $\sim 0$ near the
horizon.  There are non-vanishing stresses inside the ISCO and the
``stress edge'' is near the horizon.  These stresses
appear due to organized magnetic flux, but evidently do not cause significant
transport of angular momentum (Fig.~\ref{angmmtm}) and are not expected
to be a significant source of dissipation and radiation.

Another way of evaluating the degree of coupling is to consider, by
analogy with steady state flows, the locations of ``critical points,''
defined as the radii at which various volume-averaged outgoing
characteristic speeds vanish in the coordinate frame.  For our GRMHD
simulation, the fast magnetosonic radius is located at
$r_{\rm fast} \sim 5.1M$ showing that, even for this relatively thin
accretion disk, magnetic fields do enhance the communication between
the disk and the plunging region.  On the other hand, the Alfv\'en
critical point is much further out at $8.0M$.
Since Alfv\'en waves play an important role in angular
momentum transport, this suggests that the shear coupling via magnetic
fields is weak, unlike the suggestion of \citet{krolik99} and
\citet{gammie99}.

For fitting disk models to observations the most useful quantity is
the rate of dissipation of energy as a function of radius, since this
is what determines the radiation emitted by the accretion disk.  As a
simple estimate, consider the standard theory of viscous disks in
which the energy dissipation rate is equal to the local shear stress
multiplied by $rd\Omega_K/dr$ (this is not necessarily valid for an
MHD fluid, and is even less valid in the plunging region).  In
contrast to the standard model, which assumes a vanishing torque at
the ISCO, we find $\bar{\ell}_{\rm out} =0.022\bar{\ell}_{\rm in}$ at
$r=r_{\rm ISCO}$.  This modification to the inner boundary condition
will cause the luminosity of the disk outside the ISCO to increase by
about 4\%.

\begin{figure}
\includegraphics[width=3.3in,clip]{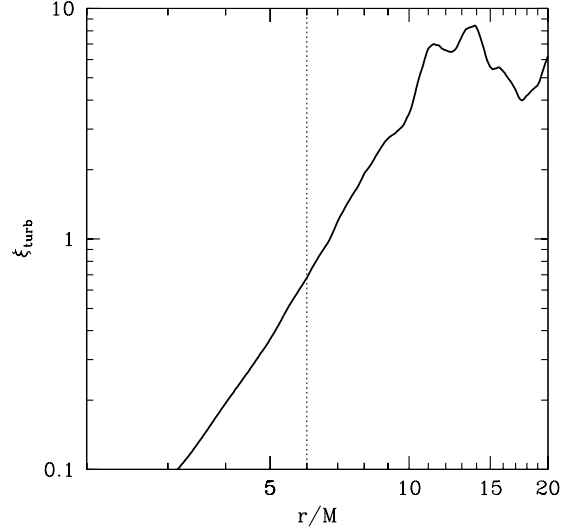} 
\caption{Time-averaged profile of $\xi_{\rm turb}$, which measures the
relative magnitude of turbulent fluctuations in the accreting gas.
The fluid becomes mostly laminar inside the ISCO.}
\label{turbulence} \end{figure}

Figure~\ref{turbulence} shows one measure of the magnitude of
turbulent fluctuations,
\begin{equation}
\xi_{\rm turb} = \frac{\sqrt{\langle(v^r)^2\rangle -\langle v^r\rangle^2}}{\langle v^r\rangle},
\end{equation}
time and angle-averaged as for Fig.~\ref{angmmtm}.  We see that the
gas at radii $\geq 10M$ is highly turbulent due to the MRI. However,
the gas flow is fairly smooth at smaller radii, suggesting that there
is little dissipation of turbulent kinetic energy inside the ISCO.

\newpage
\section{Conclusion}

For a geometrically thin accretion disk with $h/r \sim0.05-0.1$
(Fig. 1) around a non-spinning BH, we find that the specific angular
momentum profile $\bar{\ell}_{\rm in}(r)$ of the accreting magnetized gas is
quite similar to that assumed in the idealized model of NT73 (Fig. 2).
Specifically, $\bar{\ell}_{\rm in}$ is slightly smaller than, but nearly
equal to, the Keplerian value of $\bar{\ell}$ for radii outside the ISCO,
and is almost independent of $r$ inside the ISCO.  At the BH horizon,
the value of $\bar{\ell}_{\rm in}$ deviates from the NT73 value by only
2.0\%.

This result suggests that the NT73 model is a good approximation for
thin disks.  However, fast magnetosonic waves from radii as small as
$5.1M$ can escape the plunging region and communicate magnetic
stresses between the disk and the plunging regions, which means that
the zero-torque condition assumed in the NT model is not perfectly
valid.  Nevertheless, the normalized outward flux of angular momentum
$\bar{\ell}_{\rm out}$ at the ISCO is only $\sim2.2\%$ of the inward
flux $\bar{\ell}_{in}$.  Thus, the magnetic coupling across the ISCO
is quantitatively weak.  As a result, we estimate the luminosity of
the disk to be enhanced by no more than a few percent.

We find that turbulent activity is pronounced at radii beyond about
$10M$, and that the flow is nearly laminar inside the ISCO.  This
suggests that the bulk of the dissipation occurs outside the ISCO in
the disk proper, not in the plunging region. In our next paper we plan
to discuss the energy dissipation profile of 3D GRMHD simulations,
which is ultimately what determines the observed radiation.  Future
studies should also investigate the dependence of these results on
$h/r$, magnetic field geometry, black hole spin
(e.g. \citealt{gsm04}), cooling, resistivity and viscosity,
and mass distribution.

The results reported here are in qualitative agreement with those
obtained by \citet{rb08}.  They carried out a 3D non-relativistic MHD
simulation in a pseudo-Newtonian potential, whereas ours is a fully
relativistic GRMHD simulation.  Also, their initial torus had its
inner radius at the ISCO, $r_{\rm in}=6M$, whereas our torus initially
had $r_{\rm in}=20M$.  This might explain some differences in our
results, e.g., their turbulence edge is at $5.5M$ whereas ours is at
$6.8M$ (compare their Fig.~3 with our Fig.~4).  Indeed, we find that
our results are sensitive to the initial conditions if we place the
inner edge of the torus much inside $r_{\rm in}\sim 20M$.  Despite
these differences the two simulations agree on the main result, viz.,
for a geometrically thin disk, the ISCO does behave like a physical
boundary separating the disk proper from the plunging region.
\citet{beckwith08b} suggest a larger deviation from NT73,
but they considered a torus with a larger value of $h/r$
and did not use a self-consistent dissipation model.
A study of the actual dissipation in the simulation, the
goal of our next paper, will hopefully resolve these
differences.

\acknowledgements

We thank Niayesh Afshordi for stimulating discussions and useful
suggestions.  The simulations described in this paper were run on the
BlueGene/L system at the Harvard SEAS CyberInfrastructures Lab.  This
work was supported in part by NASA grant NNH07ZDA001N and NSF grant
AST-0805832.  JCM was supported by NASA's Chandra
Postdoctoral Fellowship PF7-80048.

\end{document}